\pdfoutput=1

\documentclass[11pt]{article}

\usepackage[preprint]{acl}

\usepackage{times}
\usepackage{latexsym}

\usepackage[T1]{fontenc}

\usepackage[utf8]{inputenc}

\usepackage{microtype}

\usepackage{inconsolata}

\usepackage{graphicx}

\usepackage{booktabs}
\usepackage{tabularx}
\usepackage{titlesec}
\usepackage{amstext}
\usepackage{amsmath}
\usepackage{enumitem}
\usepackage{longtable}

\newcolumntype{Z}{>{\raggedright\let\newline\\\arraybackslash\hspace{0pt}}X} 

\titlespacing*{\section}{0pt}{*0.5}{*1} 
\titlespacing*{\subsection}{0pt}{*0.5}{*1} 

\title{Humans are more gullible than LLMs \\ in believing common psychological myths}

\author{Bevan Koopman \\
  CSIRO \& The University of Queensland \\
  Brisbane, Australia \\
  \texttt{bevan.koopman@csiro.au} \\\And
  Guido Zuccon \\
  The University of Queensland \\
  Brisbane, Australia \\
  \texttt{g.zuccon@uq.edu.au} \\}

\begin{document}
\maketitle
\begin{abstract}
Despite widespread debunking, many psychological myths remain deeply entrenched. This paper investigates whether Large Language Models (LLMs) mimic human behaviour of myth belief and explores methods to mitigate such tendencies. Using 50 popular psychological myths, we evaluate myth belief across multiple LLMs under different prompting strategies, including retrieval-augmented generation and swaying prompts. Results show that LLMs exhibit significantly lower myth belief rates than humans, though user prompting can influence responses. RAG proves effective in reducing myth belief and reveals latent debiasing potential within LLMs. Our findings contribute to the emerging field of Machine Psychology and highlight how cognitive science methods can inform the evaluation and development of LLM-based systems.
\end{abstract}

\section{Introduction}

Consider the following statements:
\textit{People are either left-brained or right-brained}; 
\textit{Handwriting reveals our personality traits}; 
\textit{The polygraph (i.e., Lie Detector) test accurately detects dishonesty}.

All these are myths. They are taken from \citet{Lilienfeld:2009aa}'s \textit{50 great myths of popular psychology: Shattering widespread misconceptions about human behavior}. Despite the fact that these myths are debunked in the psychological literature, many of them are still widely believed~\cite{Meinz:2024aa} and found online~\cite{Lilienfeld:2009aa}. 

Large Language Models (LLMs) trained with vast quantities of Internet natural language data would likely encounter both content touting these myths and content refuting them. How then would an LLM respond to such myths? Would we expect the LLMs to exhibit the same myths belief patterns of people, given the online data used for training? Or is an LLM able to discern fact from fiction in its training data?

This paper aims to systematically evaluate how LLMs behave when presented with common psychological myths. The purpose of our study is both to understand how close LLMs are to human behaviour but also to consider how to mitigate myth belief. 
We pose the following research questions:
    \textbf{RQ1} Do LLMs mimic similar myth believing patterns of humans? \\
    \textbf{RQ2} How can LLMs myth belief be mitigated? \\
    \textbf{RQ3} Can a user's pre-existing bias in prompting influence LLM myth belief?

More broadly too, this paper aims to contribute to an emerging body of research on ``Machine Psychology''~\cite{Hagendorff:2024aa}, which aims to use theory and practice from human psychology to better understand LLM behaviour.

\section{Related Work}

The seminal book by~\citeauthor{Lilienfeld:2009aa} details 50 widespread myths in psychology~\cite{Lilienfeld:2009aa}. It explains where each myth came from and why it persists, while drawing on scientific evidence to debunk it. Following on from this important initial work, \citet{Furnham:2014aa} empirically evaluated belief on these 50 myths. Their study revealed that 43\% of myths were believed when evaluated with 829 human subjects. Their conclusion was that myth belief was abundant and persistent. They also noted that many widely believed myths were potentially harmful or at least socially divisive. Somewhat surprising was that education level (including some who were psychology students) did not really influence belief.

The study by \citet{Furnham:2014aa} was then replicated by \citet{Meinz:2024aa} who also tried to tease out predictors of myth belief based on participant factors such as education, cognitive ability, and personality in 150 psychology students. On education, myth belief was slightly higher for junior students but not by much. Those lower on cognitive tests believed myths more. Personality trait was also a strong predictor of myth belief with participants found to exhibit a tendency to seek knowledge less likely to believe myths. The overall human judgements from this study were made public. These, therefore, can be used as a human baseline for our study in understanding how they compare with LLMs.

The research area of ``Machine Psychology''~\cite{Hagendorff:2024aa} aims to use theory and practice from human psychology to better understand LLM behaviour. Existing work has compared whether LLMs exhibit the same cognitive errors as humans in common critical reasoning tasks \cite{Hagendorff:2023aa}. The results showed that earlier LLMs exhibit errors as humans, but later, larger models and chain-of-thought reasoning capability largely reduced any errors by LLMs.

LLMs trained on human text might exhibit the same cognitive biases of humans~\cite{Sumita:2025aa}. Empirically, some studies have shown that LLMs behave similar to humans \cite{Lampinen:2024aa,Shaki:2023aa,Suri:2024aa,Jones:2022aa}, while others show LLMs behave differently~\cite{Macmillan-Scott:2024aa,Opedal:2024aa}. Debiasing strategies generally target individual cognitive biases and are mostly implemented by prompting the LLM with a statement to avoid a specific bias.

\section{Methods}

\subsection{50 Myths Data \& Evaluation}

We utilise the 50  myths taken from the previous studies on myth belief with human subjects~\cite{Meinz:2024aa,Furnham:2014aa,Lilienfeld:2009aa}. These are provided in Table~\ref{tbl:myths}.

Previous studies with human subjects simply asked if they believed each myth statement, with the correct answer being false. Thus the evaluation and ground truth is simple: a binary label of False for each. An overall measure of myth belief \% is: $100 * \frac{1}{N} \sum_{i=1}^{N} B_i$
where $B_i$ is 1 if myth $i$ is believed, otherwise 0; $N=50$ myths.

\subsection{Prompting LLMs with Myths}

An LLM is prompted with myths according to the general prompt shown in Table~\ref{tbl:prompts}. The system prompt remains the same for all experiments with the aim of soliciting a simple True or False answer from the LLM, as well as an explanation following. An LLM response of True represents the myth was believed; False represents the myth was not believed. For some cases the model did not provide a clear True/False -- these were recorded as "Mixed Belief" in our evaluation.

Three different user prompts are evaluated. The neutral prompt is directly prompting the LLM with the text of the myth. The other two prompts aim to answer RQ2 - Can a user's pre-existing bias in prompting influence LLM myth belief. In particular, the prompt is prepended with either the text ``I'm very sceptical of'' or ``I truly believe''. This is done to understand to what degree the LLM can be swayed in its myth belief based on a particular pre-existing bias of the person asking the question. 

\begin{table}
\small
  \begin{tabularx}{1\columnwidth}{@{}l|Z@{}}
     \toprule
    System prompt &  \small \texttt{Answer if the following statement is true or false. Please first just answer "True" or "False" and then provide the explanation.} \\
    \midrule
    User Prompts & \\
    - Neutral & \small \texttt{<myth>}\\
    - Very sceptical & \small \texttt{I'm very sceptical of <myth> }\\
    - Truely believe & \small \texttt{I truly believe <myth>} \\
    - RAG Prompt & \small \texttt{Help me answer the question: <myth> given the following information: <document>} \\
    - Sway Prompts: \\
    \hspace{4pt} - Sceptic & \small \texttt{Tell my why <myth> is a myth.} \\
    \hspace{4pt} - Believer & \small \texttt{Write me a short passage explaining why my statement is true. Do not mention information about why it might be false or a myth. My statement is: <myth>.} \\
    \bottomrule
  \end{tabularx}
    \vspace{-5pt}
  \caption{Different prompts used to evaluate myth belief in LLMs. Different  prompts are used to sway the LLM in myth belief.}
  \vspace{-15pt}
  \label{tbl:prompts}
\end{table}


\subsection{Mitigation via RAG}

We implement a RAG pipeline to determine its impact on myth belief. We used MSMARCO v1 passage snapshot, which is a prebuilt index provided as part of the Pyserini IR toolkit\footnote{\url{https://github.com/castorini/pyserini/blob/master/docs/experiments-msmarco-passage.md}}, and Pyserini's BM25 implementation for retrieval. 

The 50 myth queries were issued to the retrieval system and top $k=1$ passages returned. These passages were then submitted to the LLM according to the RAG prompt  (Table~\ref{tbl:prompts}). The same three User Prompts (Neutral, Very Sceptical, Truely Believe) were also used with RAG. 

\subsection{Swaying Myth Beliefs with RAG}

RAG is used as a mitigation strategy. However, we also propose to use RAG as a means to investigate if it's possible to sway the LLM according to a pre-existing bias toward myth belief or scepticism.

To achieve this, we first prompt an LLM with the "Sway" prompts of Table~\ref{tbl:prompts}. The response is then treated as a RAG document and fed to the RAG prompt. The same is done for the "Believer" Sway prompt of Table~\ref{tbl:prompts}\footnote{For 8 myths the LLM refused to provide a passage supporting the myth. For these, we manually altered the prompt until we were able to obtain a statement support the myth.}. These two different prompts are used to induce a response from the LLM that aligns with a prior believe in the myth, and determine if that prior belief is encoded in the LLM. In addition, it helps to understand how sensitive RAG is to being fed information from different standpoints.

\section{Results \& Analysis}

\begin{figure}
  \includegraphics[width=1.01\columnwidth]{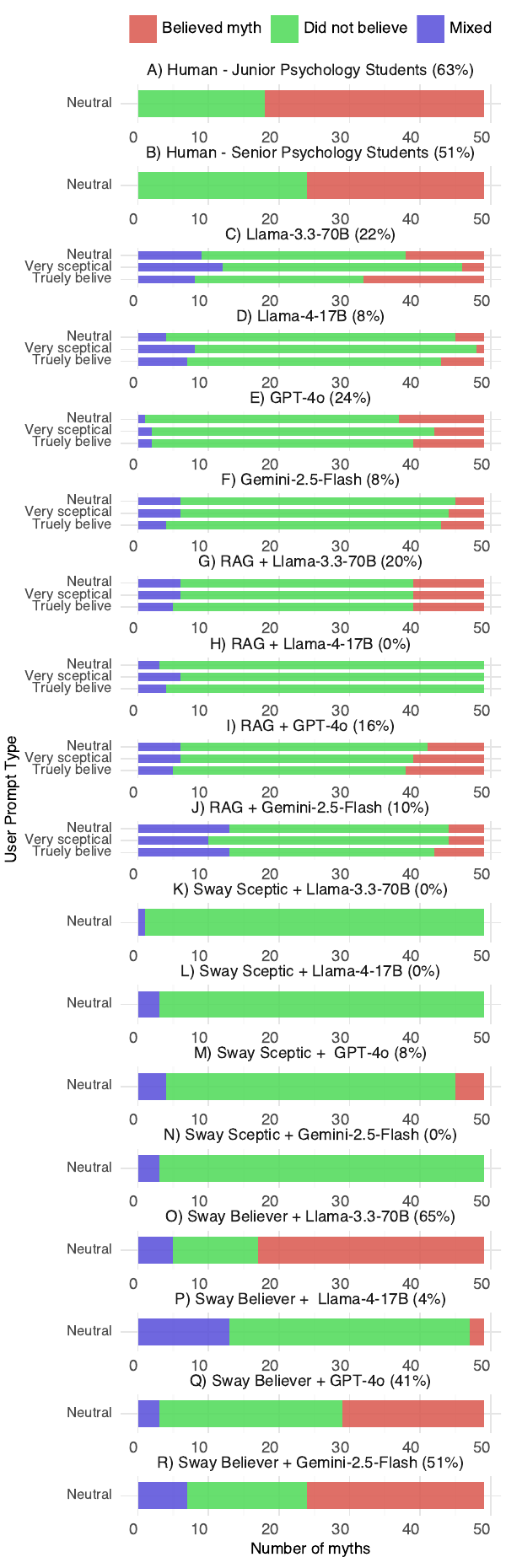}\vspace{-10pt}
  \caption{Myth belief results for different settings. Percentages are Believed Myth on Neural prompt.}
  \label{fig:results}
\end{figure}

We combine all the results for our RQs into Figure~\ref{fig:results}, detailing different aspects according to the research questions below.

\subsection{RQ1 --- Do LLMs mimic similar myth believing patterns of humans?}

The top two sub plots (A and B) of the figure show the human myth belief results from~\citet{Meinz:2024aa}. Again, these highlight the fact that myth belief is widespread (and potentially harmful). 

Sub plots C, D, E, F show the myth believe of four LLMs under the  prompt types of Neutral, Very Sceptical, Truely believe. Considering just the Neutral prompt, Gemini-2.5-Flash and LLama-4-17B exhibited only 8\% myth belief; GPT-4o exhibited 24\% and Llama-3.3-70B exhibited 22\%. This is well below the human results of 63\% and 51\%. 

There was considerable variation according to the different prompt types. In general, the Very Sceptical prompt  induced the least myth belief. The Truely Belief prompt exhibited the highest degree of myth belief, as might be expected. This highlights that a user with bias towards a particular belief can influence model responses in the way they frame their question/prompt.

\subsection{RQ2 --- Can myth belief be mitigated?}

The results of myth mitigation using RAG are shown in sub plots G, H, I and J. RAG generally reduced myth belief. For Llama 4, myth believe went to 0\% (from 8\% of subplot B). However, RAG is not a silver bullet: some models showed only a small reduction in myth belief and for Gemini (subplot J) RAG actually increased belief from 8\% (subplot F) to 10\%. These differences in comparing the RAG results and the RQ1 base setting show that different LLMs have different behaviour in their parameterised knowledge vs. in context knowledge.

The RAG setting seems far less sensitive to the different prompt types. Overall, we conclude that RAG is an effective mitigation strategy. The lack of sensitivity to prompt might also indicate that RAG is less susceptible to users bias in the way they frame their prompt/question.

In our experiments, retrieval is taken from MSMARCO, representing a general web-based document collection; the type of documents users (and RAG) might observe in general web browsing and search. Users may obviously choose more specific sources. For example, they might seek out more reputable sources (e.g., Wikipedia), which may debunk common myths. Or they might use social media where myths typically circulate. A RAG pipeline that mimics this behaviour (e.g., indexing Wikipedia vs social media posts) may exhibit different results. This is left to future work.

\subsection{RQ3 --- Can a user's pre-existing bias in prompting influence LLM myth belief?}
\vspace{-3pt}

This aimed to sway the LLM by specifically giving information that refutes or supports a myth; this was done using the RAG pipeline. As reminder, we first prompt the LLM with one of the two sway prompts (Sceptic and Believer from Table~\ref{tbl:prompts}), and then use the response as the document in our RAG pipeline. The Sway Sceptic approach (sub plots K, L, M and N) resulted in very low myth belief (only GPT-4o had a non zero myth believe of 8\% in subplot M). These results show that the LLM did in fact have internal knowledge to refute myths, if it could be extracted by specific prompting and injected into a RAG pipeline. 

Considering the Sway Believer setting (sub plot O, P, Q, R), we observe that the LLM was swayed considerably --- 65\% for Llama 3, 51\% for Gemini and 41\% for GPT4. Llama 4 behaved quite differently with only 4\%. Mixed belief was also higher. On manual inspection, Mixed Belief responses often contained a statement indicating that while the myth might be true, there was no scientific evidence to indicate so. These cases showed some contention between the knowledge passed in RAG statement supporting the myth (contextual knowledge) and the model prior stance (parametrised knowledge), which was observed in RQ1. 

\section{Discussion \& Conclusion}

The experimental results show that LLMs exhibit far lower myth belief behaviour than humans. It's difficult to determine the underlying reason why. It could be that in training, the model simply observed more information refuting myths than supporting them. Unfortunately, most LLMs today do not divulge the data used in training. Alternatively, low myth belief may actually be an emergent property of LLMs. A controlled experiment of pre-training an LLM with specific training data is left to future work to help answer these questions.

The results raise the issue of the interplay between parameter knowledge (coming from training model hyper-parameters) and contextual knowledge (coming from information provided in the prompt). We see that the different prompting strategies (Very Sceptical vs Truely Believe) were able to influence the model in their respective ways. This highlights that users with pre-existing biases in belief can pose their question in a way to confirm their bias. 
%
%
This could lead to increased echo chambers, siloing or conspiracy theorising. We advocate for mitigation strategies to prevent such harms.

We noted that the RAG pipelines were far less susceptible to user bias in prompts. Here the contextualised knowledge (prompt) contained both the bias question but also retrieval results; the retrieval results helped to overcome the biased framing of the question. This shows that there is not just an interplay between parameterised knowledge and contextual knowledge, but also an interplay within contextual knowledge that influences LLM behaviour. 
%
%
We can conclude that RAG might be a fruitful way to mitigate user bias and myth belief.

The Sway (Sceptic and Believer) prompts were designed to sway the LLM in a specific direction. The fact that the model could produce statements strongly refuting (Sceptic) and strongly supporting (Believer) the myths showed models parameters do encode both these points of view. This finding may have wider implications for the field of Machine Psychology as it is an example of models divergence from human behaviour, whereby humans typically maintain just one point of view.

LLMs are becoming increasingly pervasive and integrated into both human-machine interaction and human decision making. In this complex human-LLM interaction environment, the interplay between human cognitive bias, LLM parameterised knowledge and LLM contextual knowledge all work to influence these interactions --- and ultimately influence human decision making. This paper is one step in better understanding this interplay under the guise of human myth belief. Our aim is to contribute a better understanding of Machine Psychology, with the ultimate goal of improving human-machine interactions.

\clearpage

\section*{Limitations}

This study showed that different prompts influenced myth belief. The prompts used (Table~\ref{tbl:prompts}) were developed to, and did indeed, exhibit different behaviour. However, we did not perform any extensive prompt engineering. A core tenant of this study is that the prompts can strongly influence LLM responses (and that a user with a particular bias can influence this response). We were able to show this with the few prompts we evaluated, but a more systematic study into how user's frame their questions would likely yield more insights.

The paper considered just one retrieval system in the RAG experiments. We did not systematically evaluate how different retrieval systems impact myth belief. This includes both considering different retrieval models as well as looking at different corpora used by the retrieval system. Different corpora, in particular, may strongly influence myth belief. For example, retrieval from a corpus of scientific literature would likely result in retrieval of articles debunking myths. In contrast, retrieval from a corpus of social media content would likely result in retrieval of posts spreading myths. A followup study can investigate the interplay between retrieval and myth belief in a controlled manner. This further study is really focused at looking at how contextual knowledge (via RAG) influence LLM behaviour.

The LLMs we considered (GPT, Llama and Gemini) are all quite similar in that they are general purpose LLMs from major vendors. Most were trained with eye for the quality of training data and contain guardrails and safety measures to reduce harmful responses. An LLM without this focus on quality training data and response quality may exhibit very different behaviour. Such an LLM might more closely exhibit the myth belief behaviour of humans. A follow up study could control the training data used for  an LLM and perform addition pre-training on different data that specifically supports and refutes myths. The fully open LLM OLMo 2~\cite{OLMo:2025aa} provides a good foundation to conduct such a study.

\bibliography{references-myths2025}

\appendix

\onecolumn

\section{50 Myths}

Table~\ref{tbl:myths} provides the 50 myths taken from \citet{Lilienfeld:2009aa}. Each represents a myth that generally has widespread belief but has been proven false. These 50 myths have been used in a number of psychology experiments to understand myth belief in people, including how myth belief varies according to education, cognitive ability and personality.

\begin{table}[h!]
\resizebox{.87\textwidth}{!}{
  \begin{tabular}{l}
  	\toprule
We only use 10\% of our brains. \\ 
Most people in their 40s and 50s experience a midlife crisis. \\ 
Keeping a positive attitude can help keep cancer at bay. \\ 
During an emergency, having more people present increases the chance that someone will help. \\ 
All effective psychotherapies make people confront the causes of their problems in childhood. \\ 
The polygraph (Lie Detector) test accurately detects dishonesty. \\ 
Hypnosis causes a “trance” state, different from wakefulness. \\ 
Dreams hold symbolic meaning. \\ 
People are either left-brained or right-brained. \\ 
Intelligence tests are biased against certain groups. \\ 
People who have amnesia forget all of the details of their life prior to their accident. \\ 
Psychiatric labels stigmatize and cause harm to people. \\ 
Handwriting reveals our personality traits. \\ 
Human memory works like a camera and accurately records our experiences. \\ 
Our eyes emit light that causes us to see. \\ 
A major cause of psychological problems is low self-esteem. \\ 
If someone confesses to a crime, they are almost always guilty of it. \\ 
Recently there has been a massive epidemic of autism in childhood. \\ 
Stress is the primary cause of ulcers. \\ 
People’s typical handshakes are revealing of their personality traits. \\ 
Reversing letters is the central characteristic of dyslexia. \\ 
Adolescence is a time of psychological chaos. \\ 
Extrasensory perception (i.e., ESP or “psychic feelings”) is a scientifically established phenomenon. \\ 
If we inherit a trait, we can’t change it. \\ 
The only effective treatment for alcoholics is abstinence. \\ 
People with Schizophrenia have multiple personalities. \\ 
Raising children similarly leads to similar adult personalities. \\ 
It’s better to let out anger than to hold it in. \\ 
Happiness mostly comes from our external circumstances. \\ 
Hypnosis can help retrieve suppressed or forgotten memories. \\ 
Most mentally ill people are violent. \\ 
Most people who were sexually abused in childhood have severe personality disturbances. \\ 
Adult children of alcoholics display distinctive symptoms. \\ 
Playing classical music to infants increases their intelligence. \\ 
Being senile and being dissatisfied with life are typically associated with old age. \\ 
Only people who are very depressed commit suicide. \\ 
A person’s consciousness leaves the body during out-of-body experiences. \\ 
If you’re unsure of an answer when taking a test, it’s best to stick with your first hunch. \\ 
Individuals are capable of learning new information while asleep. \\ 
Criminal profiling helps solve cases. \\ 
Subliminal messages can persuade us to purchase products. \\ 
Men and women communicate in completely different ways. \\ 
Electroconvulsive (shock) therapy is a physically dangerous and brutal treatment. \\ 
Students learn best when teaching styles are matched to their learning styles. \\ 
Criminals commonly use the insanity defense to get off free. \\ 
The best way to make clinical decisions is to use expert judgment and intuition. \\ 
It is common to repress the memories of traumatic events. \\ 
Psychiatric hospital admissions and crimes increase during full moons. \\ 
We are romantically attracted to people who are different from us. \\
    \bottomrule
  \end{tabular}
}
  \caption{ 50 myths taken from \citet{Lilienfeld:2009aa}.}
  \label{tbl:myths}
\end{table}

\section{Statistical Significance Tests}

Table \ref{tab:chi2_significance_matrix} reports statistical significance tests differences between models in belief rates via Chi-Squared Test. Models labels are provide in Table~\ref{tbl:model_legend}.

\begin{longtable}{llllllllllllllllll}
\caption{Chi-square significance matrix (* indicates $p < 0.05$)} \label{tab:chi2_significance_matrix} \\
\toprule
 & B & C & D & E & F & G & H & I & J & K & L & M & N & O & P & Q & R \\
\midrule
\endfirsthead
\caption[]{Chi-square significance matrix (* indicates $p < 0.05$)} \\
\toprule
 & B & C & D & E & F & G & H & I & J & K & L & M & N & O & P & Q & R \\
\midrule
\endhead
\midrule
\multicolumn{18}{r}{Continued on next page} \\
\midrule
\endfoot
\bottomrule
\endlastfoot
A & * &  &  &  &  &  &  &  &  &  &  &  &  & * &  &  &  \\
B &  &  &  & * & * & * &  &  &  &  &  &  &  & * &  &  &  \\
C &  &  &  & * & * &  &  &  &  &  &  &  &  &  &  &  &  \\
D &  & * &  & * & * &  &  &  &  & * &  &  &  &  &  &  &  \\
E &  &  &  &  &  &  &  &  &  &  &  &  &  &  &  &  &  \\
G &  & * &  & * & * &  &  &  &  &  &  &  &  & * &  &  &  \\
H &  &  &  & * & * &  &  &  &  &  &  &  &  &  & * &  &  \\
I &  & * & * & * & * & * &  &  & * &  & * &  & * &  & * & * &  \\
J &  & * & * & * & * & * &  &  &  & * &  &  & * &  & * &  &  \\
K &  &  &  &  & * &  &  &  &  &  &  &  &  &  &  &  &  \\
L &  &  &  &  & * &  &  &  &  &  &  &  &  &  & * &  &  \\
M &  &  &  &  &  &  &  &  &  &  &  &  &  &  &  &  &  \\
N &  &  &  &  &  &  &  &  &  &  &  &  &  &  & * &  &  \\
O &  &  &  &  &  &  &  &  &  &  &  &  &  &  &  &  &  \\
P &  & * & * & * & * &  &  &  &  & * &  &  &  &  &  &  &  \\
Q &  & * &  & * & * & * & * &  & * &  &  &  &  & * & * &  & * \\
R &  & * &  & * &  & * &  &  &  &  &  &  &  & * &  &  &  \\
\end{longtable}

\begin{table}[h!]
\centering
\caption{Model Legend for statistical significance results of Table~\ref{tab:chi2_significance_matrix}.}
\label{tbl:model_legend}
\begin{tabular}{cl}
\toprule
\textbf{Label} & \textbf{Model Name} \\
\midrule
A & Human - Junior Psychology Students (63\%) \\
B & Human - Senior Psychology Students (51\%) \\
C & Llama-3.3-70B (22\%) \\
D & Llama-4-17B (8\%) \\
E & GPT-4o (24\%) \\
F & Gemini-2.5-Flash (8\%) \\
G & RAG + Llama-3.3-70B (20\%) \\
H & RAG + Llama-4-17B (0\%) \\
I & RAG + GPT-4o (16\%) \\
J & RAG + Gemini-2.5-Flash (10\%) \\
K & Sway Sceptic + Llama-3.3-70B (0\%) \\
L & Sway Sceptic + Llama-4-17B (0\%) \\
M & Sway Sceptic + GPT-4o (8\%) \\
N & Sway Sceptic + Gemini-2.5-Flash (0\%) \\
O & Sway Believer + Llama-3.3-70B (65\%) \\
P & Sway Believer + Llama-4-17B (4\%) \\
Q & Sway Believer + GPT-4o (41\%) \\
R & Sway Believer + Gemini-2.5-Flash (51\%) \\
\bottomrule
\end{tabular}
\end{table}

\end{document}